# NEURAL NETWORK MODELING OF DATA WITH GAPS: METHOD OF PRINCIPAL CURVES, CARLEMAN's FORMULA, AND OTHER

**A.N. Gorban, A.A. Rossiev**

Institute of Computational Modeling SD RAS,
Akademgorodok, Krasnoyarsk-36, 660036, Russian Federation
E-mail: gorban@cc.krascience.rssi.ru

**D. C. Wunsch II**

Applied Computational Intelligence Laboratory
Department of Electrical Engineering, Texas Tech University,
Lubbock, TX 79409-3102, USA
E-mail: Dwunsch@aol.com

A method of modeling data with gaps by a sequence of curves has been developed. The new method is a generalization of iterative construction of singular expansion of matrices with gaps.

Under discussion are three versions of the method featuring clear physical interpretation:

1) linear – modeling the data by a sequence of linear manifolds of small dimension;

2) quasilinear – constructing "principal curves: (or "principal surfaces"), univalently projected on the linear principal components;

3) essentially non-linear – based on constructing "principal curves": (principal strings and beams) employing the variation principle; the iteration implementation of this method is close to Kohonen self-organizing maps.

The derived dependencies are extrapolated by Carleman's formulas. The method is interpreted as a construction of neural network conveyor designed to solve the following problems:

1) to fill gaps in data;

2) to repair data – to correct initial data values in such a way as to make the constructed models work best;

3) to construct a calculator to fill gaps in the data line fed to the input.

**Introduction**

Information about an object under study often carries gaps and erroneous values. Generally, data need preliminary processing, e.g. to fill gaps and correct distorted values. Explicit or implicit such a preprocessing takes place almost every time.

Let there be a table of data ($a_{ij}$), its rows correspond to objects, and its columns correspond to features. Then, let a part of information in the table be missing – there are gaps (some $a_{ij}=@$, symbol @ is used to denote gaps in the data). The main problem that arises in this connection is to fill the existing gaps plausibly. There exists an associated problem – to "repair" the table: to distinguish the data with unfeasible values and correct them. In addition, it is useful to construct a calculator associated with the table, to fill the gaps in the incoming data about new objects and repair these new data (under the assumption that the data are interrelated in the same way as in the initial table). Such a calculator implies these data about new objects to be interrelated as in the initial table.

It should be emphasized that these problems deny discussion of either true data values or statistical provability, they discuss plausibility only.

The described problems are especially difficult (and simultaneously – attractive) in the cases when the density of gaps is high, their location is irregular, and the amount of data is small, for instance, the number of rows is approximately equal to the number of columns.

Ordinary regression algorithms – construction of empirical dependences of some data on other – are inapplicable here. If the gaps are located irregularly, this would require to construct dependencies of unknown data with respect to all their known locations in the table. This would actually mean to construct $2^{n-1}$ dependences, where $n$ is the number of features. To recover any unknown set of data at least something should be known. In this connection one needs to use the method of modeling data by manifolds of small dimension.

The essence of the method is as follows. The vector of data $x$ with $k$ gaps is represented as the $k$-dimensional linear manifold $L_x$, parallel to $k$ coordinate axes corresponding to the missing data.

Under a priori restrictions on the missing values instead of $L_x$ we have a rectangular parallelepiped $P_x \subset L_x$. The manifold $M$ of a given small dimension (in most cases – a curve) approximating the data in the best way and satisfying certain regularity conditions is searched for. For the complete vectors of data the accuracy of approximation is determined as a regular distance from a point to a set (the lower bound of the distances to the points of the set). For the incomplete data in its stead use is made of the lower bound of the distances between the points of $M$ and $L_x$ (or, accordingly, $P_x$). From the data the closest to them points of $M$ are subtracted – we obtain a residual, and the process is repeated until the residuals are close enough to zero.

Proximity of the linear manifold $L_x$ or parallelepiped $P_x$ to zero means that the distance from zero to the point of $L_x$ (accordingly, $P_x$) closest to it is small. To further specify the method is to determine how the manifold $M$ is constructed.

The idea of modeling data by manifolds of small dimension has been conceived long ago.

Its most widespread, old and feasible implementation for the data without gaps is the classical method of principal components. The method calls for modeling the data by their orthogonal projections over "principal components" - eigen vectors of the correlation matrix with corresponding largest eigen values. Another algebraic interpretation of the principal component method is singular expansion of the data table. Generally, to present data with sufficient accuracy requires relatively few principal components and dimension is reduced by a factor of tens.

Its generalization for the nonlinear case – the method of principal curves – was proposed in 1988. There are generalizations of the classical method of principal components for data with gaps as well.

The work describes a technique of constructing a system of models for incomplete data. In the simplest case they are a generalization of the classical (linear) method of principal components for data with gaps. The quasilinear method below is superstructed over the linear and employs its results. Finally, the formalism of self-organizing curves is employed to construct an essentially non-linear method.

Each method is accompanied with a physical interpretation illustrating similarity of the methods and their sequential development.

In the general case the method of modeling data with gaps by manifolds (linear and non-linear) of small dimension appears to be more efficient as compared to ordinary regression equations.

The algorithms being developed can be applied when the data matrix cannot be reduced by rearrangement of lines and columns to the following box-diagonal form:

$$A = \begin{bmatrix} A_1 & @ & ... & @ \\ @ & A_2 & ... & @ \\ ... & ... & ... & ... \\ @ & @ & ... & A_n \end{bmatrix},$$

where @ are the rectangular matrices with unknown elements. To establish connections between different $A_i$ boxes in such tables is impossible.

**Statement of the Problem**

Let there be a rectangular table $A=(a_{ij})$ the cells of which are filled with real numbers or symbol @ denoting absence of data.

It is required to construct models that make possible to solve the following three problems related to recovery of the missing data:

1. To fill the gaps in data plausibly.

2. To repair the data, i.e. to correct their values in such a way as to make the models constructed work best.

3. To construct by the table available a calculator that would fill the gaps in data and repair them as they arrive (assuming the data in the line arriving at the input to be connected by the same relations as in the initial table).

This raises the question: how (*in which metric*) to evaluate the error of the model? To choose the measure of the error is required both to construct the models and to test them. From the viewpoint of simplicity of calculations the most attractive is the least squares method (LSM). By this method the error is calculated as the sum of squares of deviations over all known data (Mean Square Error – MSE). Yet, arbitrary rule associated with the choice of scale, i.e. normalization of data is present here, too.

The classical method of principal components generally normalizes the initial data to unit dispersion. After such a normalization the first principal component is defined as such a direction (vector) that has maximum dispersion of orthogonal projections of data on it. It corresponds to the main axis of the ellipsoid of concentration.

More comprehensively, the principal component method (PCM) implements transition to the new coordinate system $y_1,...,y_p$ in the initial space of features $x_1,...,x_p$. These new coordinates are constructed as a system of orthonormalized linear combinations:

$$\begin{cases} y_j(x) = w_{1j}(x_1 - m_1) + \ldots + w_{pj}(x_p - m_p); \\ \sum_{i=1}^{p} w_{ij}^2 = 1, (j = 1..p); \\ \sum_{i=1}^{p} w_{ij} w_{jk} = 0, (j, k = 1..p, j \neq k), \end{cases} \quad (1.1)$$

where $m_i$ is the mathematical expectation of feature $x_i$.

The coefficients are chosen for the first principal component $y_1(x)$ to have the highest dispersion among all possible linear normalized combinations of linear features of (1.1) type. Geometrically it appears as a new coordinate system $y_1$ oriented along the main axis of the ellipsoid of dispersion of the objects in the study sampling in the space of features $x_1,\ldots,x_p$. The second principal component features the highest dispersion among all remaining linear transformations of form (1.1) not correlated with the first principal component. It is interpreted as the highest prolateness of the dispersion ellipsoid perpendicular to the first principal component. The principal components to follow are defined in analogy.

Normalization to unit dispersion is not always consistent with the essence of the matter. In addition to the mean square deviation $\sigma$ of a given value the function of natural scale is claimed by accuracy of its measurement and, even more important, its variation allowance.

The notion of "allowance" originates from technical applications and denotes the arbitrary rule in the magnitude of values that can be allowed without sacrificing the solution of the problem. Allowance is defined by individual users or special agreements about standard allowances. It is the allowance value that is most probably the best natural measurement scale. It should be borne in mind, however, that this value is determined not by the data table only, but also by the problems to be solved by it.

In the initial construction the number of principal components is equal to the number of initial features – this is merely a transition to a new coordinate system. Still, there is no necessity to calculate all principal components and even more so to retain them in the model. Suffice is to retain several of them. Taking from $p$ data $m$ principal components ($m<p$) we arrive at the so-called $m$-factor model.

The basic model of factor analysis can be written with the following system of equalities:

$$x_i = \sum_{j=1}^{m} l_{ij} f_j + \varepsilon_i, \quad i=1..p, \; m<p. \quad (1.2)$$

It is assumed that the value of each feature $x_i$ can be expressed by weighted sum of latent variables (simple factors) $f_j$, their number being smaller than the number of initial features, and the residual error $\varepsilon_i$ with dispersion $\sigma^2(\varepsilon_i)$ affecting $x_i$ only. This residual error is also called a specific factor.

In the simplest model of factor analysis it is thought that the factors $f_j$ are mutually independent and their dispersions are equal to unit, and the random values $\varepsilon_i$ are also independent of each other and of some factor $f_j$. Most commonly the factors $f_j$ are chosen proportional to the first $m$ principal components. The accuracy of the model is evaluated by residual dispersions $\sigma^2(\varepsilon_i)$ and the sum of these values for all features.

Everywhere below assume the data to be normalized in a reasonable way (e.g. with appropriate allowances) and evaluate errors by the least squares method.

# 1. The Linear Model

## 1.1. Singular Expansion of Tables with Gaps

The material of this section is not immediately employed to process data, yet it provides a simplest example and a prototype for further constructions.

Let there be a rectangular table $A=(a_{ij})$ the cells of which are filled with real numbers or symbol @ denoting absence of data.

The problem is to approximate $A$ best by a matrix of form $x_i y_j$ by the least squares method.

$$\Phi = \sum_{\substack{i,j \\ a_{ij} \neq @}} (a_{ij} - x_i y_j)^2 \to \min.$$

The problem is solved by successive iterations by explicit formulas. With the vector $y_j$ fixed the values $x_i$ providing minimum to the form (1.3) univalently and simply can be derived from equalities $\partial\Phi/\partial x_i = 0$:

$$x_i = \left(\sum_{\substack{j \\ a_{ij} \neq @}} a_{ij} y_j\right) \bigg/ \left(\sum_{\substack{j \\ a_{ij} \neq @}} (y_j)^2\right).$$

Analogously, with the vector $x_i$ fixed, the value $y_j$ providing minimum to form (1.3) is explicitly determined from equalities $\partial\Phi/\partial y_j = 0$:

$$y_j = \left(\sum_{\substack{i \\ a_{ij} \neq @}} a_{ij} x_i\right) \bigg/ \left(\sum_{\substack{i \\ a_{ij} \neq @}} (x_i)^2\right).$$

The initial values are:

$y$ is random normalized to 1 (i.e. $\sum_j y_j^2 = 1$).

The halting criterion is the smallness of relative improvement $\Delta\Phi/\Phi$, where $\Delta\Phi$ is the decrease of value $\Phi$ obtained in one cycle, and $\Phi$ is the current value. The second criterion is the smallness of the value $\Phi$ per se. The procedure comes to a halt when $\Delta\Phi/\Phi<\varepsilon$ or $\Phi<\delta$ for certain $\varepsilon, \delta<0$. As a result for the given matrix $A$ find the best approximation by the matrix $P_1$ of the form $x_i y_j$. Further on, look for $A-P_1$ for the best approximation of the same form $P_2$ and so on, until, e.g. the norm A does not sufficiently approach zero.

Thus, the initial matrix $A$ is presented in the form of a sum of matrices of rank 1, i.e. $A=P_1+P_2+\ldots+P_q$.

## 1.2. Geometrical Interpretation

Each line of the matrix is the vector of data $a$ with $k$ gaps, presented as the $k$–dimensional manifold $L_a$, parallel to $k$ coordinate axes corresponding to the missing data. With a priori constraints on the missing values the location of $L_a$ is taken by the rectangular parallelepiped $P_a \subset L_a$.

The vector $y$ is, at this, a straight line crossing the origin of coordinates and in the best way (in a certain precise sense) approximating the initial data, and, thus, while the vector $x$ represents a set of projections of initial data over the obtained straight line.

As a result we have an iteration process of modeling data. Its essence is in the fact that the best (in a certain precise sense) model – linear space $M$ of small dimension - is constructed for the initial data. Then subtract from the data $a$ ($L_a$ or $P_a$, accordingly) projections $Pr_M(a)$ and have deviations from the first model. Construct a simple model for this set of deviations and so on, until all deviations are sufficiently close to zero.

**1.3. Principal Component Method for Tables with Gaps**

Following the method described in the previous section derive straight lines crossing the origin of coordinates. Such homogeneous models are not always required. Expand the initial table not over the matrices of form $P=x_iy_j$ but over the matrices of form $P=x_iy_j+b_j$. This brings us to the next problem:

$$\Phi = \sum_{\substack{i,j \\ a_{ij} \neq @}} (a_{ij} - x_i y_j - b_j)^2 \to \min. \tag{1.4}$$

Solution of the problem (1.4) yields models of data by the linear manifolds not necessarily crossing the origin of coordinates.

The basic procedure is to find the best approximation of the table with gaps by matrix of form $x_iy_j+b_j$. [4,5].

With the vectors $y_j$ and $b_j$ fixed the values $x_i$, providing minimum to the form (1.4) unambiguously and simply are defined from equalities $\partial\Phi/\partial x_i=0$:

$$x_i = \left(\sum_{\substack{j \\ a_{ij} \neq @}} (a_{ij} - b_j) y_j \right) \Bigg/ \left(\sum_{\substack{j \\ a_{ij} \neq @}} (y_j)^2 \right).$$

In analogy, with the vector $x_i$ fixed the values $y_j$ и $b_j$, providing minimum to the form (1.4) are explicitly defined from two equalities $\partial\Phi/\partial y_j=0$ and $\partial\Phi/\partial b_j=0$:

$$\begin{cases} y_j A_{01}^j + b_j A_{00}^j = B_0^j \\ y_j A_{11}^j + b_j A_{10}^j = B_1^j \end{cases}, \text{ where } A_{kl}^j = \sum_{\substack{i \\ a_{ij} \neq @}} x_i^{k+l}, \; B_k^j = \sum_{\substack{i \\ a_{ij} \neq @}} a_{ij} x_i^k, \; k=0..1, \; l=0..1.$$

Expressing $b_j$ from the first equation and substituting the derived value into the second we have:

$$y_j = \frac{B_1^j - B_0^j \frac{A_{10}^j}{A_{00}^j}}{A_{11}^j - A_{01}^j \frac{A_{10}^j}{A_{00}^j}}, \; b_j = \frac{B_0^j - y_j A_{01}^j}{A_{00}^j}.$$

Initial values:

$y$ is random, normalized to 1 (i.e. $\sum_j y_j^2 = 1$) $b_j = \frac{1}{n_j} \sum_{\substack{i \\ a_{ij} \neq @}} a_{ij}$, where $n_j = \sum_{\substack{i \\ a_{ij} \neq @}} 1$ (number of known data), i.e. $b_j$ is determined as mean value in a column.

Setting practically arbitrary initial approximations for $y_j$ and $b_j$, find value $x_i$ then assuming $y_j$ and $b_j$ unknown find their values for fixed $x_i$ and so on – these simple iterations converge.

As for the problem (1.3) the halting criterion is the smallness of relative improvement $\Delta\Phi/\Phi$,, where $\Delta\Phi$ is the decrease of the value $\Phi$ obtained in a cycle, and $\Phi$ is the current value per se.

The second criterion is the smallness of the value $\Phi$ as it is. The procedure halts when $\Delta\Phi/\Phi<\varepsilon$ or $\Phi<\delta$ for some $\varepsilon, \delta<0$.

*Successive exhaustion of the matrix A.* Find for a given matrix $A$ the best approximation by the matrix $P_1$ of the form $x_iy_j+b_j$. Then find for $A-P_1$ the best approximation of the same form $P_2$ and so on. The checking can be done, e.g. by the residual dispersion of columns.

In the case of no gaps the method described brings to ordinary principal components – singular expansion of centered initial table of data. In this case, starting with $q=2$, $P_q = x_i^q y_j^q$ ($b=0$). In the general case this is not necessarily so. *It should be emphasized that centering (transition to zero mean values) is not applicable to the data with gaps.*

*Q-factor filling of gaps* is their definition from the sum of $Q$ of obtained matrices of the form $x_iy_j+b_j$,

*Q-factor repairing of a table* is its substitution with the Q of obtained matrices of the form $x_iy_j+b_j$.

Describe the procedure of recovery of data in the line $a_j$ with gaps (some $a_j=@$) arriving for processing. Let there be constructed a sequence of matrices $P_q$ of the form $x_iy_j+b_j$ ($P_q = x_i^q y_j^q + b_j^q$), exhausting the initial matrix $A$ with a preset accuracy. For each $q$ determine the number $x^q(a)$ and vector $a_j^q$ with a given line:

$$a_j^0 = a_j \ (a_j \neq @);$$

$$x^q(a) = \left(\sum_{\substack{j \\ a_j \neq @}}(a_j^{q-1} - b_j^q)y_j^q\right) \Big/ \left(\sum_{\substack{j \\ a_j \neq @}}(y_j^q)^2\right); \quad (1.5)$$

$$a_j^q = a_j^{q-1} - b_j^q - x^q(a)y_j^q, \ (a_j \neq @);$$

here the manifold M is a straight line, the coordinates of points on M are assigned by the parametric equation $z_j=ty_j+b_j$ and the projection $\Pr_M(a)$ is defined by (2):

$\Pr(a)=t(a)y_j+b_j$;

$$t(a) = \left(\sum_{\substack{j \\ a_j \neq @}}(a_j - b_j)y_j\right) \Big/ \left(\sum_{\substack{j \\ a_j \neq @}}(y_j)^2\right). \quad (1.6)$$

For recovery of the $Q$-factor assume

$$\bar{a}_j = \sum_{q=1}^{Q} x^q(a)y_j^q + b_j^q, \ (a_j \neq @). \quad (1.7)$$

When there are no gaps the derived straight lines are orthogonal and we have an orthogonal system of factors. For the incomplete data this is not the case, but the derived system of factors can be orthogonalized, the process consists in recovery of the initial table by the derived system of factors following which the system is again recalculated, now on complemented data.

### 1.4. Mechanical Interpretation

Consider mechanical concepts underlying construction of the modeling manifold.

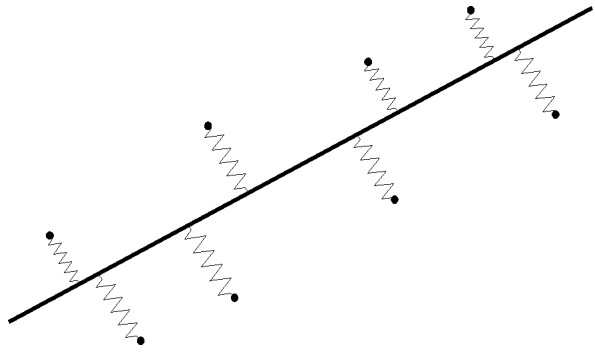

Fig. 1

Let the space of data have a straight rigid beam (Fig. 1). Let every datum be connected with the beam by a spring with the end of the spring moving along the beam. Fix the initial position of the beam and find such a position of springs that corresponds to the minimum of elastic energy. Then fix the position of spring ends on the beam, release the beam and let it regain mechanical equilibrium. Then fix the beam in a new position and again release the spring ends.

The system monotonically approaches equilibrium – minimum of its energy, as at every stage the strain energy decreases. The obtained beam is nothing but the first principal component. The projection of the datum is, at this, determined by the site where the spring corresponding to it is fixed on the beam.

The iteration finding of the coefficients $x_i$, $y_j$ and $b_j$ described in the previous section is fully analogous to the described mechanical process: the position of the beam is determined by the coefficients $y_j$, $b_j$ and the points where the springs are fixed – by the coefficients $x_i$.

The data with gaps are modeled by hard rods (one gap), planes, etc. A corresponding spring can freely move along these objects. This means that the spring efficiently connects the shadow (projection) of the beam on the subspace of known data with the point in this subspace.

### 1.5. Two-Dimensional and Three-Dimensional Models

In place of the linear manifold of dimension 1 it is possible to take linear manifolds of dimension 2 or 3. Calculations, at this, are done by the absolutely analogous formulas:

$$\Phi = \sum_{\substack{i,j \\ a_{ij} \neq @}} (a_{ij} - x_{1i}y_{1j} - x_{2i}y_{2j} - b_j)^2 \to \min, \quad (1.8)$$

$$\Phi = \sum_{\substack{i,j \\ a_{ij} \neq @}} (a_{ij} - x_{1i}y_{1j} - x_{2i}y_{2j} - x_{3i}y_{3j} - b_j)^2 \to \min, \quad (1.9)$$

Multidimensional linear models do not produce anything of substantial novelty. Increase of dimension of linear models can be efficient in quasilinear models only.

## 2. The Quasilinear Model

### 2.1. The Principal Curve Method

Take the principal curve method as a prototype to construct the nonlinear models [1-3].

*Definition 1*: (Principal curve). Let $X$ denote a random vector in the subspace $R^d$ and let there be $n$ data vectors. The principal curve $f \subset R^d$ is a smooth ($C^\infty$) curve (of dimension 1) in $R^d$ parametrized by the parameter $\lambda \in A \subset R^1$, going through the middle of $d$ –dimensional data described by $X$,

$$f(\lambda) = \begin{bmatrix} f_1(\lambda) \\ \vdots \\ f_d(\lambda) \end{bmatrix} = E\{X | \lambda_f(X) = \lambda\}, \qquad (2.1)$$

where

$$\lambda_f(x) = \sup_\lambda \{\lambda : \|x - f(\lambda)\| = \inf_\mu \|x - f(\mu)\|\} \qquad (2.2)$$

is the function of projection on the curve.

In brief, every point on the principal curve is mean value of all data points projected on it. The mean square error between the data points and their closest projections on the principal curve in the $j$-th iteration is as follows:

$$MSE^{(j)} = E\left\{\left\|X - f^{(j)}[\lambda_{f^{(j)}}(X)]\right\|^2\right\}. \qquad (2.3)$$

The algorithm of constructing the principal curve is iteration involving calculation of mathematical expectation (2.1) and projection at each iteration. Tolerance (TOL) at the $j$-th iteration is defined as relative alteration of MSE as follows:

$$TOL^{(j)} = \frac{|MSE^{(j)} - MSE^{(j-1)}|}{|MSE^{(j-1)}|}. \qquad (2.4)$$

Construction is interrupted when TOL is below the set threshold.

The principal curve can be open or closed. To calculate a closed curve we add a segment involving extreme points.

## 2.2. A Method of Constructing Quasilinear Models

A simplest version of nonlinear factor analysis is superstructed over the linear one. It is proposed to use quasilinear models admitting simple explicit formulas for data processing based on the described algorithm of constructing linear models.

Let, as in the case of linear models, there be a table with gaps $A$. Quasilinear model is constructed in several stages:

1) To construct a linear model is to solve the problem (1.4). For the sake of certainty assume $(y,b)=0$, $(y,y)=1$ – this can be achieved always.

2) Interpolate (smoothen). Construct a vector-function $f(t)$ minimizing the functional

$$\Phi = \sum_{\substack{i,j \\ a_{ij} \neq @}} (a_{ij} - f_j(\sum_k a_{ik} y_k))^2 + \alpha \int_{-\infty}^{+\infty} (f''(t))^2 dt, \qquad (2.5)$$

where $\alpha > 0$ is the smoothening parameter.

The problem is solved by cubic splines whose coefficients are found from the equality to zero of corresponding partial derivatives $\Phi$ (2.5) on a certain uniform network. For those nodes of the network where the data do not arrive, the coefficients are found from the consistency conditions (the continuity of the function and the continuity of the first and second derivatives).

The problem can also be solved with a polynomial of a small degree, however, even though such a solution makes possible to obtain a satisfactory interpolation with small calculation efforts, it cannot yield good extrapolation.

3) Extrapolation. The obtained function is extrapolated over the entire real axis.

### 2.2.1. Interpolation with a polynomial of small degree

In the case of approximating with a polynomial of degree $n$ the problem is to best approximate the matrix $A$ by the polynomials of the form $f_j(x) = f_n^j x^n + f_{n-1}^j x^{n-1} + ... + f_1^j x + f_0^j$, i.e.:

$$\Phi = \sum_{\substack{i,j \\ a_{ij} \neq @}} (a_{ij} - f_j(\sum_k a_{ik} y_k))^2 + \alpha \int_{-\infty}^{+\infty} (f''(t))^2 dt \to \min,$$ where $\alpha > 0$ – is the smoothening parameter.

As the values $x_i = (a_i, y)$, where $a_i$ is the $i$-th line of the matrix $A$, are fixed (calculated at the previous stage) the values of the coefficients of the polynomial $f_k^j$ ($k=0..n$) providing the minimum to the functional $\Phi$ are determined from the system of equalities $\partial \Phi / \partial f_k^j = 0$ ($k=0..n$) as follows:

$$f_j'(t) = \sum_{k=1}^{n} k f_k^j x^{k-1}, \quad f_j''(t) = \sum_{k=2}^{n} k(k-1) f_k^j x^{k-2}, \text{ then:}$$

$$\left( f_j''(t) \right)^2 = \sum_{k,l=2}^{n} k(k-1) l(l-1) f_k^j f_l^j x^{k+l-4}.$$

With account of the fact that the function sought for has been defined at the segment [-1,1]:

$$\int_{-1}^{+1} \left( f_j''(t) \right)^2 dt = \sum_{k,l=2}^{n} f_k^j f_l^j k(k-1) l(l-1) \frac{x^{k+l-3}}{k+l-3} \bigg|_{-1}^{+1} = \sum_{\substack{k,l=2 \\ k+l \text{ - четное}}}^{n} f_k^j f_l^j k(k-1) l(l-1) \frac{2}{k+l-3}.$$

Then:

$$\frac{d}{df_k^j} \int_{-1}^{+1} \left( f_j''(t) \right)^2 dt = \sum_{\substack{l=2 \\ l \neq k \\ k+l \text{ - четное}}}^{n} f_l^j k(k-1) l(l-1) \frac{2}{k+l-3} + 2 f_k^j k^2 (k-1)^2 \frac{2}{2k-3}$$

Finally, for $k=0..n$ we have

$$\frac{\partial \Phi}{\partial f_k^j} = -2 \sum_{\substack{i \\ a_{ij} \neq @}} (a_{ij} - \sum_{l=0}^{n} f_l^j x_i^l) x_i^k +$$

$$+ \sum_{\substack{l=2 \\ l \neq k \\ k+l \text{ - четное}}}^{n} f_l^j k(k-1) l(l-1) \frac{2}{k+l-3} + 2 f_k^j k^2 (k-1)^2 \frac{2}{2k-3} = 0$$

Grouping the coefficients at $f_k^j$ ($k=0..n$) we have

$$\sum_{l=0}^{n} A_{kl}^j(0) f_l^j = B_k^j(0), \text{ for } k=0..n, \text{ where:}$$

$$A_{kl}^{j}(0) = \sum_{\substack{i \\ a_{ij} \neq @}} x_i^{k+l} + \begin{cases} 0 & \text{, when } k < 2 \text{ or } l < 2 \text{ or } k+l \text{ is odd} \\ 2k^2(k-1)^2 \dfrac{2}{2k-3} & \text{, when } k = l \\ k(k-1)l(l-1)\dfrac{2}{k+l-3} & \text{, when } k \neq l \end{cases}$$

$$B_k^j(0) = \sum_{\substack{i \\ a_{ij} \neq @}} a_{ij} x_i^k.$$

Having solved the obtained system of equations have the following recurrent formulas to find $f_k^j$ (k=0..n):

$$A_{kl}^j(0) = \sum_{\substack{i \\ a_{ij} \neq @}} x_i^{k+l}, \quad B_k^j(0) = \sum_{\substack{i \\ a_{ij} \neq @}} a_{ij} x_i^k,$$

$$A_{kl}^j(m+1) = A_{kl}^j(m) - A_{ml}^j(m)\frac{A_{km}^j(m)}{A_{mm}^j(m)}, \quad B_k^j(m+1) = B_k^j(m) - B_m^j(m)\frac{A_{km}^j(m)}{A_{mm}^j(m)},$$

$$f_k^j = \frac{B_k^j(k) - \sum_{l=k+1}^n A_{kl}^j(k) f_l^j}{A_{kk}^j(k)}, \text{ for } k=n..0.$$

### 2.2.2. Interpolation by cubic splines

The following smoothening problem is to be solved by cubic splines:

$$\Phi = \sum_{\substack{i,j \\ a_{ij} \neq @}} (a_{ij} - f_j(\sum_k a_{ik} y_k))^2 + \alpha \int_{-\infty}^{+\infty} (f''(t))^2 dt \to \min,$$

where $\alpha > 0$ is the smoothening parameter.

Let there be a segment [-1,1]. Divide it into $n$ parts by points $\bar{x}_s$, s=0..n where $-1 = \bar{x}_0 < \bar{x}_1 < ... < \bar{x}_{n-1} < \bar{x}_n = 1$.

At this: $h_s = \bar{x}_s - \bar{x}_{s-1}$.

Let $\bar{x} = x - \bar{x}_{s-1}$, where $\bar{x}_{s-1} < x < \bar{x}_s$, s=1..n, then have:

$$f(x) = f_{s3}\bar{x}^3 + f_{s2}\bar{x}^2 + f_{s1}\bar{x} + f_{s0} = \sum_{l=0}^{3} f_{sl}\bar{x}^l.$$

Calculate the derivatives:

$$f'(x) = \sum_{l=1}^{3} l f_{sl}\bar{x}^l, \quad f''(x) = \sum_{l=2}^{3} l(l-1) f_{sl}\bar{x}^l.$$

The value of the integral with the smoothness coefficient is:

$$I = \int_{-1}^{1} (f''(t))^2 dt = \sum_{s=1}^{n-1} \int_{\bar{x}_{s-1}}^{\bar{x}_s} \left( \sum_{l=2}^{3} l(l-1) f_{sl} t \right)^2 dt = \sum_{s=1}^{n-1} \left( \sum_{k,l=2}^{3} f_{sk} f_{sl} k(k-1) l(l-1) \frac{t^{k+l-3}}{k+l-3} \bigg|_{\bar{x}_{s-1}}^{\bar{x}_s} \right) =$$

$$= \sum_{s=1}^{n-1} \left( \sum_{k,l=2}^{3} f_{sk} f_{sl} k(k-1) l(l-1) \frac{h_s^{k+l-3}}{k+l-3} \right)$$

Thus, the initial smoothening problem is written in the following form:

$$\Phi = H + \alpha I \to \min,$$

where $H = \sum_{\substack{i \\ a_i \neq @}} \left( a_i - \sum_{l=0}^{3} f_{sl} \bar{x}_i^l \right)$, $I = \sum_{s=1}^{n-1} \left( \sum_{k,l=2}^{3} f_{sk} f_{sl} k(k-1) l(l-1) \frac{h_s^{k+l-3}}{k+l-3} \right)$.

The derived spline function must be continuous at the network nodes together with its first and second derivatives:

$$\begin{cases} f(\bar{x}_s - 0) = f(\bar{x}_s + 0) \\ f'(\bar{x}_s - 0) = f'(\bar{x}_s + 0) \\ f''(\bar{x}_s - 0) = f''(\bar{x}_s + 0) \end{cases}, \text{ where } s = 1..n-1.$$

Substituting respective values, have conditions for the coefficients of splines:

$$\begin{cases} f_{s+1,0} = f_{s3} h_s^3 + f_{s2} h_s^2 + f_{s1} h_s + f_{s0} \\ f_{s+1,1} = 3 f_{s3} h_s^2 + 2 f_{s2} h_s + f_{s1} \\ f_{s+1,2} = 3 f_{s3} h_s + f_{s2} \end{cases}, \text{ where } s = 1..n-1.$$

The initial smoothening problem is solved by the method of Lagrange multipliers:

$$L = \Phi + \sum_{s=1}^{n-1} \lambda_{s0} \varphi_{s0} + \sum_{s=1}^{n-1} \lambda_{s1} \varphi_{s1} + \sum_{s=1}^{n-1} \lambda_{s2} \varphi_{s2},$$

where:

$$\begin{cases} \varphi_{s0} = f_{s+1,0} - f_{s3} h_s^3 - f_{s2} h_s^2 - f_{s1} h_s - f_{s0} \\ \varphi_{s1} = f_{s+1,1} - 3 f_{s3} h_s^2 - 2 f_{s2} h_s - f_{s1} \\ \varphi_{s2} = f_{s+1,2} - 3 f_{s3} h_s - f_{s2} \end{cases}, \text{ for } s = 1..n$$

are linear restrictions derived from the continuity condition.

The following system is to be solved:

$$\begin{cases} \dfrac{\partial L}{\partial f_{sl}} = 0, \text{ for } l = 0..3, s = 1..n \\ \dfrac{\partial L}{\partial \lambda_{sl}} = 0, \text{ for } l = 0..2, s = 1..n-1 \end{cases},$$

i.e. the system of $(7n-3)$ equations and with $(7n-3)$ unknowns.

The first lines of this system can be written as follows:

$$\frac{\partial L}{\partial f_{sl}} = \frac{\partial H}{\partial f_{sl}} + \alpha \frac{\partial I}{\partial f_{sl}} + L_{sl}, \text{ where:}$$

$$\frac{\partial H}{\partial f_{sl}} = -2\sum_{\substack{i \\ a_i \neq @}} \left( a_i - \sum_{p=0}^{3} f_{sp}\bar{x}_i^p \right) \bar{x}_i^l , \quad \frac{\partial I}{\partial f_{sl}} = \sum_{\substack{k=2 \\ k \neq l}}^{3} f_{sk} k(k-1) l(l-1) \frac{2}{k+l-3} + 2 f_{sl} l^2 (l-1)^2 \frac{2}{2l-3},$$

$$\begin{cases} L_{s0} = \lambda_{s-1,0} - \lambda_{s0} \\ L_{s1} = \lambda_{s-1,1} - \lambda_{s0} h_s - \lambda_{s1} \\ L_{s2} = \lambda_{s-1,2} - \lambda_{s0} h_s^2 - \lambda_{s1} 2h_s - \lambda_{s2} \\ L_{s3} = -\lambda_{s0} h_s^3 - \lambda_{s1} 3h_s^2 - \lambda_{s2} 3h_s \end{cases},$$

the coefficients with index $s-1$ are, at this, taken for $s=2..n$, the coefficients with index s are taken for $s=1..n-1$.

Substituting all this into the initial system one can see that its form is actually box-diagonal, where the size of individual boxes is sufficiently small, therefore, in spite of its large size $((7n-3)\times(7n-3))$ it can be solved with good accuracy even by simple numerical methods.

## 2.3. Extrapolation and Carleman's Formulas

### 2.3.1. The Problem of Extrapolation, Optimum Analytical Continuation and Carleman's Formula

The problem of extrapolating the available data beyond the limits of dispersion is well known. Its solution cannot be omitted – there is no guarantee that the data to follow will get exactly into the variation range of those available and it is not possible always limit the scope beyond this range. The necessity to construct new formulas for all possible data values is called by two circumstances more: first – the smoothened dependence constructed at the first stage is fundamentally interpolation and cannot be extrapolated, second – actually it carries in explicit form the information about each line of the data matrix. Smoothening, e.g. by merely a polynomial of a small degree by the least squares method is free from the second drawback (the information is "reduced" to several coefficients), but does not yield a good extrapolation.

Common (but fairly rough) extrapolation by the straight lines is: the resulting function $f(t)$ is extrapolated from the segment (e.g. $[a,b]$) over the entire real axis owing to the first approximation constructed at the ends of the segment $f(t)=f(a)+f'(a)(t-a)$ with $t<a$ и $f(t)=f(b)+f'(b)(t-b)$ with $t>b$.

Optimal extrapolation is of more interest. Its more strict statement calls for involvment of the problem of analytical extension of the function (from the finite set of points over a straight line or space). It is also convenient to pass over from the real variable $t$ to a band on the plane of complex variables.

So, under consideration is the problem of analytical extension of the function, set on an infinite sequence of points $\{t_k\}$ ($k=1,2,\ldots$). It is required to construct a formula of extension from a *finite* set best in the following sense: the sequence of functions $f_n(t)$ obtained by extension from the set $\{t_k\}$ ($k=1,2,\ldots n$), converges faster than for all other formulas of this class. Of course, it takes to additionally define the following concepts: what are "formulas of this class", what convergence is meant, etc. All this has been done in relevant mathematical literature [6].

The smoothened vector-function $f(t)$ can be optimally extrapolated from a certain finite set $\{t_k\}$ (not necessarily related with projections over the straight line $z_j=ty_j+b_j$ of initial lines of data)

$$f(t) \approx ty + b + \sum_{k=1}^{m}(f(t_k) - t_k y - b)\frac{2(e^{\lambda t} - e^{\lambda t_k})}{\lambda(e^{\lambda t} + e^{\lambda t_k})(t - t_k)}\prod_{\substack{j=1 \\ j \neq k}}^{m}\frac{(e^{\lambda t_k} + e^{\lambda t_j})(e^{\lambda t} - e^{\lambda t_j})}{(e^{\lambda t_k} - e^{\lambda t_j})(e^{\lambda t} + e^{\lambda t_j})}, \quad (2.6)$$

where $\lambda$ is the parameter of the methods specifying the width of the band on the plane of complex numbers, where the extrapolated function is holomorphic with a guarantee (the width is $\pi/\lambda$).

Generally, to extrapolate according to Carleman for a set of nodes $\{t_k\}$ we take points uniformly placed on the segment, but not the initial experimental data. The values of $f(t)$ at these points we find by the interpolation formulas.

By Carleman's formulas we extrapolate the deviation of the curve $f(t)$ from the straight line $ty+b$. The Carleman's formulas provide for good extrapolation of analytical functions over the entire straight line (it cannot be guaranteed, of course, that in every specific case it is the formula (2.6) that will assure the best extrapolation, yet, there are several theorems that the formula (2.6) and relevant formulas yield the best approximation for different classes of analytical functions [6]).

Smoothing and extrapolation by Carleman's formula can be combined into a single process, i.e. interpolation and extrapolation can be done simultaneously. This is possible when interpolation is done with the following version of the Carleman's formula:

$$f(t) \approx ty + b + \sum_{k=1}^{m} f_k \cdot \frac{2(e^{\lambda t} - e^{\lambda t_k})}{\lambda(e^{\lambda t} + e^{\lambda t_k})(t - t_k)}\prod_{\substack{j=1 \\ j \neq k}}^{m}\frac{(e^{\lambda t_k} + e^{\lambda t_j})(e^{\lambda t} - e^{\lambda t_j})}{(e^{\lambda t_k} - e^{\lambda t_j})(e^{\lambda t} + e^{\lambda t_j})}$$

where the coefficients $f_k$ can be found from the formula (2.5) in analogy with the smoothening problem by cubic splines.

### 2.3.2. Interpolation and Optimal Smoothening by the Carleman's Formula

The smoothening problem to be solved with the Carleman's formula is as follows:

$\Phi = H + \alpha I \to \min$, where:

$$H = \sum_{\substack{i,j \\ a_{ij} \neq @}} (a_{ij} - f_j(\sum_k a_{ik} y_k))^2, \quad I = \int_{-\infty}^{+\infty}(f''(t))^2 dt$$

$f(t) = ty + b + \sum_{k=1}^{m} f_k E_k(t)$ is the vector function,

$$E_k(t) = \frac{2(e^{\lambda t} - e^{\lambda t_k})}{\lambda(e^{\lambda t} + e^{\lambda t_k})(t - t_k)} \cdot \prod_{\substack{j=1 \\ j \neq k}}^{m}\frac{(e^{\lambda t_k} + e^{\lambda t_j})(e^{\lambda t} - e^{\lambda t_j})}{(e^{\lambda t_k} + e^{\lambda t_j})(e^{\lambda t} + e^{\lambda t_j})}$$

and, finally, $\alpha > 0$ is the smoothening parameter.

(Since the calculations for each $j$ are done in the same manner, for the calculations they can be neglected).

The values of coefficients $f_k$ ($k=0..n$), providing the minimum to the functional $\Phi$, are determined from the system of equalities $\partial \Phi / \partial f_k = 0$ ($k=0..n$) as follows:

$$\frac{\partial H}{\partial f_k} = -2\sum_i \left(a_i - t_i y - b - \sum_{l=1}^{m} f_l E_l(t_i)\right) E_k(t_i) = 0,$$

$$2\sum_{l=1}^{m} f_l \left( \sum_i E_l(t_i) E_k(t_i) \right) = 2\sum_i (a_i - t_i y - b) E_k(t_i).$$

So, not taking into consideration the smoothness integral we have the following system of linear equation with respect to $f_k$, which can be solved numerically

$$\sum_{l=1}^{m} A_{kl} f_l = B_k,$$

where $A_{kl} = 2\sum_i E_k(t_i) E_l(t_i)$, $B_k = 2\sum_i (a_i - t_i y - b) E_k(t_i)$.

The value of the smoothness integral $I$ is:

$$I = \int_{-\infty}^{+\infty} \left( \sum_{k,l=1}^{m} f_k f_l E_k''(t) E_l''(t) \right) dt = \sum_{k,l=1}^{m} \left( \int_{-\infty}^{+\infty} f_k f_l E_k''(t) E_l''(t) dt \right) =$$

$$= \sum_{k,l=1}^{m} f_k f_l \left( \int_{-\infty}^{+\infty} E_k''(t) E_l''(t) dt \right) = \sum_{k,l=1}^{m} f_k f_l I_{kl}, \text{ где } I_{kl} = \int_{-\infty}^{+\infty} E_k''(t) E_l''(t) dt.$$

$$\frac{\partial I}{\partial f_k} = \sum_{l=1}^{m} f_l I_{kl}^*, \text{ where } I_{kl}^* = \begin{cases} I_{kl}, l \neq k, \\ 2I_{kl}, l = k. \end{cases}$$

With account of the smoothness integral $I$ the system of linear equation determining the point of the minimum of the functional $\Phi = H + \alpha I$ takes the form

$$\sum_{l=1}^{m} A_{kl} f_l = B_k, \qquad (**)$$

where $\quad A_{kl} = 2\sum_i E_k(t_i) E_l(t_i) + \alpha I_{kl}^*$, $B_k = 2\sum_i (a_i - t_i y - b) E_k(t_i)$.

The values of $I_{kl}$ are found by numerical integration. The system of linear equations to define $f_k$ is also solved numerically.

### 2.4. Mechanical Interpretation

Assuming the beam to be able to deviate elastically from its straight form we have the following picture (Fig. 2)

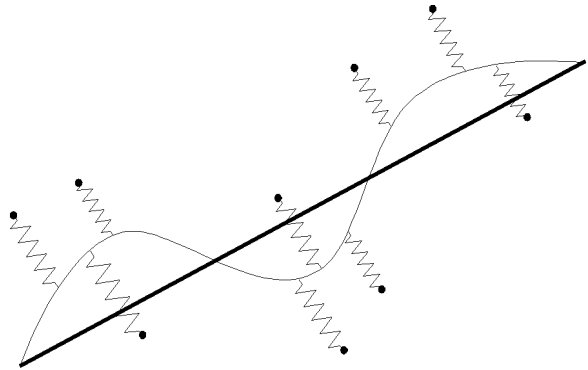

Fig. 2.

The points where the springs are fixed on the beam are determined by the projections on the straight beam (as the model is quasilinear).

The problem arises to determine behavior of the beam ends beyond the boundaries of the data range – the above described extrapolation problem.

**2.5. Application of Quasilinear Models**

A point on the constructed curve $f(t)$, corresponding to the complete ("complete") data vector a is constructed as $f((a,y))$, this is the quasilinearity of the method: First we find the projection of the data vector over the straight line $Pr(a)=ty+b$, $t=(a,y)$ after that we construct a point on the curve $f(t)$. This is also true for the incomplete data vectors – first we find on the straight line the nearest point $t(a)$, then - respective point on the curve $f(t)$ with $t=t(a)$.

After the curve $f(t)$ is constructed the data matrix is displaced with the matrix of deviation from the model. Then again we find the best approximation of the form $x_i y_j+b_j$ for the matrix of deviations, again construct smoothening, following which we extrapolate by Carleman and so on, until the deviations sufficiently approach zero.

As a result the initial table takes the form of the $Q$-factor model:

$$a_{ij} \cong \sum_q f_j^q(t_i^q). \qquad (2.7)$$

If $a_{ij} \neq @$, this formula approximates the initial data, otherwise it yields a method of data recovery.

**3. Neural Conveyor**

The constructed algorithm makes possible neural network interpretation. Corresponded to each curve $f_q(t)$ is one summator (its weights are the coordinates of the vector $y^q$), a set of $n$ free summands ("thresholds") – coordinates of the vector $b^q$ and $n$ nonlinear converters each of which calculates one coordinate of a point on the curve by the formula (2.6). Such a "neuron" acts on the vector $a$ of input signals (with gaps) as follows: $t(a)$ is calculated by the formula (1.6) (operation of the summator), then the nonlinear elements calculate $f^q(t(a))$ after that the difference $f_j^q(t(a))$ $(a_j \neq @)$ is transmitted to the following neuron. With $a$ traveling along this conveyor the sum of values of $f_j^q(t(a))$ $(a_j=@)$ builds up. It is these sums that form the vector of output signals – the proposed values of the missing data. Should the need arise to make repairs of data the sum of values of $f_j^q(t(a))$ is built up for each coordinate $j$.

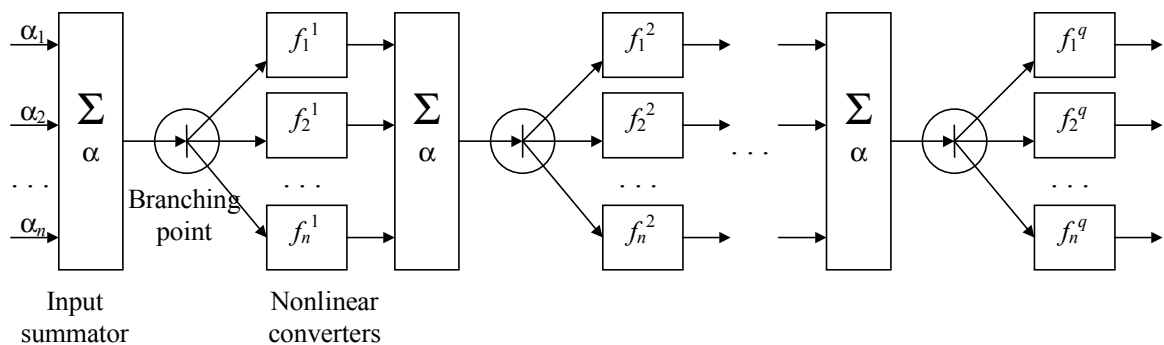

Fig. 3

The structure of neurons is not standard (Fig. 3) – it has one input summator and $n$ nonlinear converters (in compliance with the dimension of the data vector).

Operation of the summator is not quite ordinary either – for incomplete data vectors it calculates the scalar product with available data and performs additional normalization. This additional normalization of the input weights of the summator takes into account only those with known respective values of the input vector coordinates).

The described methods of constructing these neurons is noteworthy. Their characteristics are calculated in turn, the first to be constructed is, at this, the summator (solving the problem (1.4)), then – the nonlinear converters (by Carleman's formulas), after that – the summator of the following neuron and so on.

All constructed neurons work alternately (in the ordinary sense there are as many layers as there are neurons), but they form a conveyor and the released neurons can pass over to the next data vector, therefore, in successive arrival of the data the processing time is proportional to the number of neurons, but performance (the number of processed data vectors in a time unit) is determined by the actuation time of one neuron and is independent of their number.

## 4. The Self-Organizing Curves (SOC)

### 4.1. The Idea of Self-Organizing Curves

The quasilinear factors are good for the "medium nonlinear problems" where the contribution of the linear part is relatively substantial (about 50% or more). For essentially nonlinear problems it is but natural to use some approximation of principal curves instead of linear principal components.

Instead of linear manifolds of small dimension we propose to use corresponding self-organizing maps (SOM) or, in other words – self-organizing curves (SOC). The method we use is somewhat different from the method of Kohonen maps ([7] see also comprehensible presentation in [8]) by more transparent physical interpretation and explicit form of the variation principle.

Let SOC be defined by a set of points (kernels) successively placed on a curve (in the first approximation let SOC be simply a broken line $Y$) and it is required to map on it a set of points of data $X=\{x_i\}$. Introduce operator $\Pi$, which for every vector $x \in X$ associates the nearest to it point from $Y$:

$$x \xrightarrow{\Pi} y_j, \|y_j - x\|^2 \to \min, \qquad (4.1)$$

each kernel $y_j$ is associated with its taxon

$$K_j = \left\{ x \in X \middle| x \xrightarrow{\Pi} y_j \right\}. \qquad (4.2)$$

The method of constructing SOC resembles the methods of dynamic kernels except for the additional restrictions on connectivity and elasticity. The minimized value is constructed from the following summands:

the measure of data approximation -

$$D_1 = \sum_j \sum_{x \in K_j} \|x - y_j\|^2, \qquad (4.3)$$

the measure of connectivity (the points close on the curve must get close on the data subspace) -

$$D_2 = \sum_j \|y_j - y_{j+1}\|^2, \qquad (4.4)$$

the measure of nonlinearity -

$$D_3 = \sum_j \|2y_j - y_{j-1} - y_{j+1}\|^2. \qquad (4.5)$$

So, to construct SOC we have to minimize the functional:

$$D = \frac{D_1}{|X|} + \lambda \frac{D_2}{m} + \mu \frac{D_3}{m} \to \min, \qquad (4.6)$$

where $\lambda$, $\mu$ are the parameters of connectivity and nonlinearity – "moduli of elasticity" (division by the number of points $|X|$ and the number of kernels $m$ means normalization "in one summand" and makes possible to use identical methods of varying $\lambda$ and $\mu$ for samplings of different size).

When division of a set of data into taxons is fixed, SOC are constructed unambiguously – a simple linear problem is solved. When the position of kernels is also fixed the taxons can also be easily constructed by the formulas (4.1), (4.2). Dividing the problem into successive search of kernels – taxons – kernels we have an algorithm whose convergence is ensured by the fact that the D criterion decreases with its each step (4.6).

### 4.1.1. SOC Construction Algorithm

Let there be a table $(a_{ij})$ with gaps (some $a_{ij}=@$) and a set of kernels $Y=\{y^k\}$ ($k=0..m-1$). The formulas (4.1-4.5) can be rewritten as follows ($a_i$ is the $i$-th line of the table $(a_{ij})$):

$$a_i \xrightarrow{\Pi} y^k, \sum_i (a_{ij} - y_j^k)^2 \to \min, \qquad (4.1^*)$$

$$K_k = \left\{ a_i \in (a_{ij}) \,\Big|\, a_i \xrightarrow{\Pi} y^k \right\}. \qquad (4.2^*)$$

$$D_1 = \sum_{k=0}^{m-1} \sum_{i \in K_k} \sum_{\substack{j \\ a_{ij} \neq @}} (a_{ij} - y_j^k)^2, \qquad (4.3^*)$$

$$D_2 = \sum_{k=0}^{m-2} \sum_j (y_j^k - y_j^{k+1})^2, \qquad (4.4^*)$$

$$D_3 = \sum_{k=1}^{m-2} \sum_j (2y_j^k - y_j^{k-1} - y_j^{k+1})^2. \qquad (4.5^*)$$

It is required to solve the following problem:

$$D = \frac{D_1}{n} + \lambda \frac{D_2}{m} + \mu \frac{D_3}{m} \to \min, \qquad (4.6^*)$$

where $n$ is number of lines of the matrix $(a_{ij})$.

The values $y_j^k$, providing the minimum to the form (4.6*) with the given fragmentation of $K_k$, are determined from the system of equalities $\partial D / \partial y_j^k = 0$:

$$\frac{\partial D}{\partial y_j^k} = \frac{1}{n} \cdot \frac{\partial D_1}{\partial y_j^k} + \frac{\lambda}{m} \cdot \frac{\partial D_2}{\partial y_j^k} + \frac{\mu}{m} \cdot \frac{\partial D_3}{\partial y_j^k}, \qquad (4.7)$$

$$\frac{\partial D_1}{\partial y_j^k} = -2 \sum_{\substack{i \in K_k \\ a_{ij} \neq @}} (a_{ij} - y_j^k), \qquad (4.8)$$

$$\frac{\partial D_2}{\partial y_j^k} = -2(y_j^{k-1} - y_j^k) + 2(y_j^k - y_j^{k+1}),  \quad (4.9)$$

$$\frac{\partial D_3}{\partial y_j^k} = -2(2y_j^{k-1} - y_j^{k-2} - y_j^k) + 4(2y_j^k - y_j^{k-1} - y_j^{k+1}) - 2(2y_j^{k+1} - y_j^k - y_j^{k+2}). \quad (4.10)$$

For each $j$ we have a system of $k$ linear equation with respect to $y_j^k$:

$$A_{kj}^{-2} y_j^{k-2} + A_{kj}^{-1} y_j^{k-1} + A_{kj}^{0} y_j^{k} + A_{kj}^{1} y_j^{k+1} + A_{kj}^{2} y_j^{k+2} = B_j^k, \text{ where}$$

is a pentadiagonal matrix whose coefficients are determined from equations (4.7) – (4.10).

The numerical solution can be obtained by the sweep method.

### 4.2. The Local Minimum Problem

As opposed to the linear and quasilinear cases this problem of minimizing the functional (4.6.) is not convex and there arise difficulties associated with getting to the local minimum range. This can result in unsatisfactory solution of the problem.

Even though there are numerous methods to solve this problem, we would rather dwell upon the multigrid method and the "annealing" method.

### 4.3. The Annealing Method

The so-called "annealing" method is to add a certain function to the initial function being minimized, which can smoothen the local minimums to take the minimization process from the local minimum. Then the additional function approaches zero.

For our problem we propose to use the following "annealing" method:

Increase of coefficients $\lambda$, $\mu$ in (4.6) puts the system into very rigorous limits. Further on they gradually wane (respective coefficients decrease). In particular, for large $\mu$ we get segments close to the first principal component.

### 4.4. The Multigrid Method

Describe ordinary multigrid methods in an abstract algebraic form [11].

Assume that there is a sequence of spaces of limited dimension

$$M_0, M_1, \ldots, M_k$$

with inner products described as $(\cdot,\cdot)_i$ for each $M_i$. Assume also, that there are "interpolation" operators

$$I_i: M_i \rightarrow M_{i+1}, i=0,\ldots,k-1;$$

"restriction", operators

$$R_i: M_i \rightarrow M_{i-1}, i=1,\ldots,k;$$

and inversible operators

$$L_i: M_i \rightarrow M_i, i=0,\ldots,k.$$

The main purpose of these algorithms is to solve the following problem in $M_k$: given $f_k \in M_k$, required to find $v_k \in M_k$ such that

$$L_k v_k = f_k. \quad (4.11)$$

The peculiarity of multigrid algorithms is in the necessity of solving auxiliary problems at lower levels. Therefore, formulate the $MG_i$-algorithm for an approximate solution of the problem.

$$L_i v_i = f_i, \text{ где } f_i \in M_i \quad (4.12)$$

for every level $i \in [0,k]$. Begin with a certain initial approximation $w_0$ and as a result have an other approximation (expected to be closer to $v_i$), described as $w_1 = MG_i(w_0, f_i)$.

**$MG_i$-algorithm.** If $i=0$, then

$$w_1 = MG_0(w_0, f_0) = L_0^{-1} f_0,$$

the initial approximation has no meaning any more and $MG_0$ is completed. For $i>0$ the algorithm is defined as follows:

$A_1$ (pre-smoothening). Let $u_0 = w_0$ and define $u_{m_1}$ as follows

$$u_{l+1} = u_l - \mathfrak{S}_{i,l+1}(L_i u_l - f_i), \ l=0,1,\ldots,m_1-1;$$

$A_2$ (restriction). Let

$$g_{i-1} = R_i(L_i u_{m_1} - f_i);$$

$A_3$ (large-grid solution). Let $\tilde{w}_0 = 0 \in M_{i-1}$ and repeat $MG_{i-1}$-algorithm $\gamma$ times:

$$\tilde{w}_s = MG_{i-1}(\tilde{w}_{s-1}, g_{i-1}), \ s=1,\ldots,\gamma;$$

$A_4$ (correction). Let

$$y_0 = u_{m_1} - I_{i-1} \tilde{w}_\gamma;$$

$A_1$ (post-smoothening). Define $y_{m_2}$ as follows

$$y_{l+1} = y_l - \mathfrak{S}_{i,l+m_1+1}(L_i y_l - f_i), \ l=0,1,\ldots,m_2-1.$$

Finally have

$$w_1 = MG_i(w_0, f_i) = y_{m_2}$$

as a result of $MG_i$-algorithm.

$\mathfrak{S}_{i,l}$ ($l=1,\ldots,m_1+m_2$) is a certain linear operator. Its structure can be found in [11].

Three steps $A_2$–$A_4$ taken together are generally called a large-grid correction.

Complete multigrid $FMG$-algorithm to solve the problem (4.12) involving the upper level (4.12) at the final stage

**$FMG$-algorithm**

1. Let $\tilde{u}_0 = L_0^{-1} f_0$.
2. For $i=1,2,\ldots,k$ do as follows:
   2.1. Find $\tilde{u}_i = I_{i-1} \tilde{u}_{i-1}$;
   2.2. Repeat $MG_i$-algorithm $t$ times:
   $$\tilde{u}_i := MG_i(\tilde{u}_i, f_i).$$

At the $k$-th step of the algorithm we have the final result which is a certain approximation of the solution $\tilde{u}_k$ of the problem (4.11) for the level $k$.

Multigrid, or, in other words, hierarchical methods are based on the fact that in the process of calculation the step of the grid changes in some manner. E.g. in the cascade method originally large step is gradually reduced. The methods of the V-cycle and W-cycle have more complicated rules of grid step variation.

In this problem we propose to use a cascade method which originally employs a grid with a very large step and on this grid the interpolation function is constructed. As a result we have an initial and very rough approximation. Further on the grid is subject to gradual fragmentation, i.e. the size of the grid reduces stepwise. The initial approximating function is, at this, gradually corrected. The grid can be fragmented not uniformly, but depending on the accuracy required for different sites, i.e. for the entire range of data it is possible to use a large-step grid, and in places of particular interest the grid is fragmented to achieve the required accuracy.

As a result for the problem the multigrid method takes the following form:

The initially broken line consists of two points. After minimization of the functional (4.6) we have the principal straight line (analog of the earlier described linear models). Then this segment is divided into two parts, i.e. a new node is added. The functional (4.6) is minimized again. Further on each segment of the broken line is divided into two parts, the functional is minimized and so on, until the required accuracy is attained.

The version of this method that divides only those segments of the broken line which exceed the length set for the given step appears to be more economic.

### 4.5. The Smoothening

#### 4.5.1. The Smoothening Problem

The obtained broken line $\{y_j\}$ ($j=1..m$) can be smoothened by different methods, e.g. by cubic splines. This, however, brings forth certain difficulties associated with finding projections of the data over the smoothened curve, as this requires to solve algebraic equations of the 5$^{th}$ degree.

This called for construction of the initial broken line of a continuous projector. Corresponding to the ends of the broken line are values (-1) and 1, respectively, the projections of nodes onto the segment [-1, 1] are determined by the nodes of uniform (considered is the number of the nodes of the broken line only) or non-uniform (account is also taken of the distance between the nodes of the broken line) grid.

Final smoothening is done by Carleman's formulas. The broken line $\{y_j\}$ ($j=1..m$) replaces, at this, the principal component of the classical method, while the smoothening process is analogous to construction of the quasilinear model.

#### 4.5.2. Piecewise Linear Projection on the Broken Line

It is required to construct a mapping of linear manifolds $x=a_i \in (a_{ij})$ on a broken line determined by a set of apexes $\{y^k\}$ ($k=0…m-1$), i.e. each $x$ should be juxtaposed with a certain $t$ determining its projection on the broken line.

Let $y_k$ be an apex of the broken line nearest to $x$. Then:

    1.    1. $y_k$ is the extreme apex of the broken line, i.e. $k=0$ or $k=m-1$ (Fig. 4).

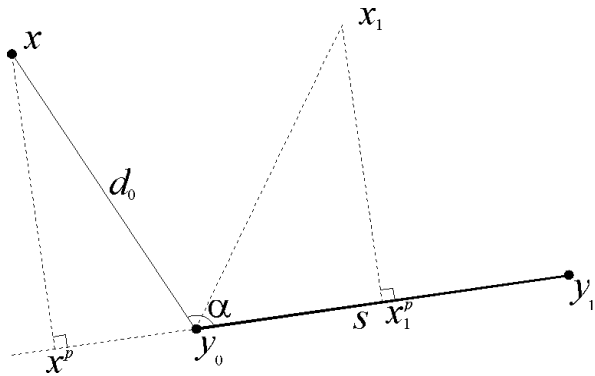

Fig. 4

Consider the operation of mapping the point $x$ on the segment of the broken line $y_0y_1$ (for $k=0$).

Let $d_0$ be the distance from $x$ to $y_0$, $\alpha$ be the angle between $y_0y_1$ and $y_0x$.

The value of the mapping function is found as a projection of $x$ on the straight line $y_0y_1$:

$$P' = \frac{2d_0 \cos\alpha}{s},$$

where $P'=0$ corresponds to the apex $y_0$, $P'=1$ is the middle of the broken line segment $y_0y_1$, and $P'<0$ is the extrapolation beyond the limits of the broken line.

For $k=m-1$ the value $P'$ is taken with the opposite sign.

2. $y_k$ is the inner apex of the broken line i.e. $0<k<m-1$ (Fig. 5).

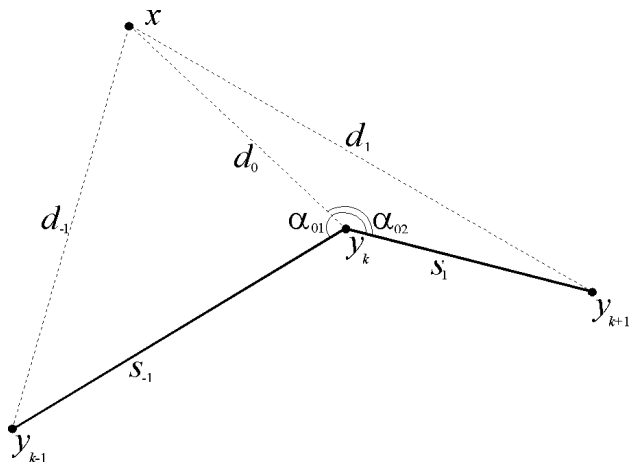 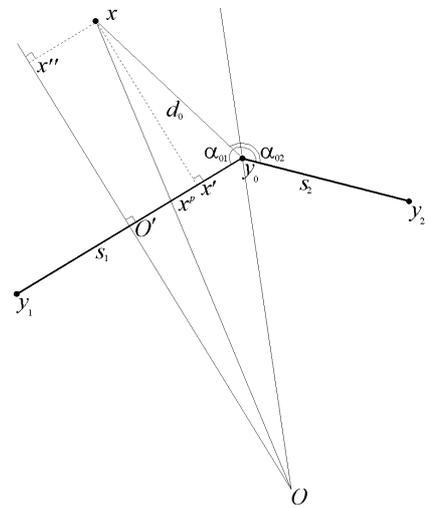

Fig. 5          Fig. 6

Let $y_{k-1}$ and $y_{k+1}$ be two neighboring apexes, $d_{-1}$, $d_0$, $d_1$ be the distances to apexes $y_{k-1}$, $y_k$, $y_{k+1}$, respectively, and $s_{-1}$ and $s_1$ be the distances between the apexes of the broken line $y_k$, $y_{k-1}$ and $y_k$, $y_{k+1}$, respectively. And let $\alpha_{01}$ be the angle between $xy_k$ and $y_ky_{k-1}$, $\alpha_{02}$ be the angle between $xy_k$ and $y_ky_{k+1}$.

If $\alpha_{01}<\alpha_{02}$, we will map onto $y_ky_{k-1}$, otherwise onto $y_ky_{k+1}$.

Consider the operation of mapping (Fig. 6) the point $x$ onto the segment of the broken line $y_0y_1$.

$$y_0O' = s_1/2, \quad y_0x' = d_0\cos\alpha_{01}, \quad x''O' = xx' = d_0\sin\alpha_{01},$$

$$xx'' = O'x' = y_0O' - y_0x' = \frac{s_1}{2} - d_0\cos\alpha_{01},$$

$$OO' = y_0O' \cdot \operatorname{tg}(\pi - \alpha_0/2) = \frac{s_1}{2}\operatorname{tg}(\pi - \alpha_0/2),$$

$$Ox'' = OO' + O'x'' = \frac{s_1}{2}\operatorname{tg}(\pi - \alpha_0/2) + d_0\sin\alpha_{01},$$

From similarity of triangles $Oxx''$ and $Oy_0O'$:

$$\frac{OO'}{Ox''} = \frac{O'x^p}{x''x},\ O'x^p = \frac{OO'}{Ox''}x''x = \frac{s_1 \cdot \operatorname{tg}(\pi - \alpha_0/2)}{s_1\operatorname{tg}(\pi - \alpha_0/2) + 2d_0\sin\alpha_{01}}\left(\frac{s_1}{2} - d_0\cos\alpha_{01}\right),$$

$$y_0x^p = O'y_0 - O'x^p = \frac{s_1}{2} - \frac{s_1 \cdot \operatorname{tg}(\pi - \alpha_0/2)}{s_1\operatorname{tg}(\pi - \alpha_0/2) + 2d_0\sin\alpha_{01}}\left(\frac{s_1}{2} - d_0\cos\alpha_{01}\right).$$

The value of the mapping function:

$$P' = 2y_0x^p/s_1 = 1 - \frac{\operatorname{tg}(\pi - \alpha_0/2)}{s_1\operatorname{tg}(\pi - \alpha_0/2) + 2d_0\sin\alpha_{01}}\left(\frac{s_1}{2} - d_0\cos\alpha_{01}\right) =$$

$$= 1 + \frac{\operatorname{tg}(\alpha_0/2)}{2d_0\sin\alpha_{01} - s_1\operatorname{tg}(\alpha_0/2)}\left(\frac{s_1}{2} - d_0\cos\alpha_{01}\right),$$

where $P'=0$ corresponds to the apex $y_0$, and $P'=1$ corresponds to the middle of $y_0y_1$.

The obtained value $P'$ is the projection with respect to one node. For the outer and inner nodes the final projection is:

$$P = -1 + \frac{2k + P'}{m - 1}.$$

### 4.6. Mechanical Interpretation

The above described linear and quasilinear models have a very strong restriction – the beam is either rigid or it is flexible, but is built along a straight line, which is essential e.g. in the case when the data are placed not along some straight line, but along a circumference (or along a strongly curved arc, at least).

To circumvent it the beam should be elastic (to be determined not by a straight line, but by a curve). This presents difficulties in defining the distance from the point to the curve in space (even more so, when it is not a point but a linear manifold).

In the above described method close to the method of Kohonen's self-organizing maps the sought for elastic beam is presented in the form of a broken line whose nodes are freely connected with the data (Fig. 7).

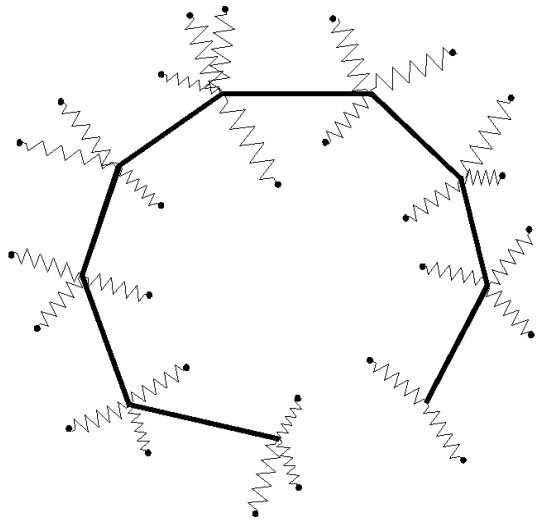

Fig. 7

In analogy to the rigid case in several iterations the system regains its equilibrium. Their number is, at this, finite, as at every step the springs decrease their total energy, and the number of possible states (methods to fix the nodes of the broken line by the springs to the data) is finite.

The introduced moduli of elasticity represent the degree of attraction of the nodes of the broken line to each other and the degree of resistance to bending at the nodes, respectively.

## 5. Experimental Results

Illustrate the process of modeling the data with gaps on the basis of the table of presidential elections in the US with 31 election situation (from 1860 to 1980). Each election in the table contains data of 12 binary features [12]:
1. Has the presidential party (P-party) been in power for more than one term? (More1)
2. Did the P-party receive more than 50% of the popular vote in the last election? (More50)
3. Was there significant activity of a third party during the election year? (Third)
4. Was there serious competition in the P-party primaries? (Conc)
5. Was the P-party candidate the president at the time of the election? (Prez)
6. Was there a depression or recession in the election year? (Depr)
7. Was there an average annual growth in the gross national product of more than 2.1% in the last term? (Val2.1)
8. Did the P-party president make any substantial political changes during his term? (Chan)
9. Did significant social tension exist during the term of the P-party? (Wave)
10. Was the P-party administration guilty of any serious mistakes or scandals? (Mist)
11. Was the P-party candidate a national hero? (R.Hero)
12. Was the O-party candidate a national hero? (O.Hero)

The table also contains information about results of elections (victory of the presidential or opposition party). The values of the binary features are equal to –1 (answer "no" for the input feature or the victory of the presidential party ) and to 1 (answer "yes" for the input feature or the victory of the opposition).

The models constructed by this table confidently predicted results of the second election of Reagan, victory of Bush over Ducacis, both victories of Clinton [13].

The degrees of approximation (in per cent of the initial value) of the table by several factors according to the model (Table 1) were as follows (Table 1):

Table 1

| № | Feature | Approximation (%) depending on the number of factors | | | | | | | | |
|---|---|---|---|---|---|---|---|---|---|---|
| | | Linear model | | | Quasiliniear model | | | SOC | | |
| | | 1 | 4 | 10 | 1 | 4 | 10 | 1 | 4 | 10 |
| 1. | *More1* | 11.88 | 59.98 | 77.36 | 25.37 | 63.75 | 95.91 | 53.49 | 80.81 | 96.85 |
| 2. | *More50* | 9.07 | 61.10 | 79.43 | 14.99 | 73.69 | 95.18 | 30.77 | 75.89 | 95.12 |
| 3. | *Third* | 29.66 | 44.89 | 91.56 | 31.73 | 66.97 | 97.45 | 32.94 | 76.83 | 96.93 |
| 4. | *Conc* | 62.30 | 63.28 | 77.84 | 69.51 | 77.12 | 90.24 | 72.63 | 78.86 | 95.42 |
| 5. | *Prez* | 31.72 | 59.68 | 80.27 | 45.58 | 68.01 | 93.03 | 56.08 | 74.85 | 95.18 |
| 6. | *Depr* | 32.17 | 52.43 | 93.38 | 37.95 | 71.08 | 95.56 | 58.86 | 80.53 | 95.62 |
| 7. | *Val2_1* | 4.12 | 37.67 | 94.19 | 6.27 | 69.22 | 96.53 | 28.80 | 72.23 | 95.44 |
| 8. | *Changes* | 2.33 | 49.87 | 86.19 | 16.81 | 61.15 | 94.95 | 13.01 | 72.01 | 93.77 |
| 9. | *Wave* | 25.13 | 62.33 | 80.34 | 33.18 | 66.82 | 95.65 | 32.68 | 63.96 | 96.71 |
| 10. | *Mist* | 50.34 | 61.05 | 86.17 | 64.83 | 70.52 | 97.55 | 60.80 | 81.35 | 96.90 |
| 11. | *R_Hero* | 33.35 | 48.12 | 90.69 | 54.86 | 66.27 | 97.52 | 27.30 | 83.67 | 95.76 |
| 12. | *O_Hero* | 36.55 | 50.07 | 92.03 | 45.69 | 68.41 | 97.55 | 52.22 | 76.42 | 96.17 |
| 13. | *Answer* | 69.22 | 69.96 | 81.78 | 92.27 | 92.85 | 96.72 | 97.82 | 98.43 | 99.50 |

If the error in the calculated value of the feature is less than 50%, this means that it is the exact value (the features are qualitative, therefore, with the error less than 50% the sign of prediction defines exact value).

For satisfactory prediction by linear models suffice are 4 factors, which indicates that this is a 4-factor problem (in the ordinary meaning of this word).

The quasilinear models, SOC-based models, in particular, usually do with only one nonlinear factor to predict satisfactorily.

The produced sets of factors were tested as follows:

> 1. A model was constructed on a complete table, then gaps were randomly added into the table, following which the procedure of filling the gaps was launched. As a result the obtained values were compared to the initial values.

The testing demonstrated that up to 25% of gaps (of the total number of initial data) were satisfactorily filled by linear and quasilinear models. For SOC-based models this index is 50%. I.e. even every other datum is eliminated from the table, still it can be recovered with satisfactory accuracy.

> 2. Gaps were randomly introduced into the table, then a model was constructed on the basis of "gap" table, following which the procedure of filling the gaps was launched. As a result the obtained values were compared to the initial values. The filling was satisfactory when the gaps amounted to 10% of the total number of initial data.

The method was actually tested on the problem of predicting complications in myocardial infarction.

The table of data on complications in myocardial infarction presents observation of 1700 patients by 126 parameters.

Experiments demonstrated that the first 15-20 quasilinear factors are sufficient to satisfactorily repair most values of features in the table. However, there are exceptions, for example "Patient's age". The calculated value frequently differed from the initial by several years for every set of initial biomedical data without direct indications of the age. This makes possible meaningful interpretation: frequently the real medical-biological age of a person is not equal to its formal age by the birth certificate.

**Discussion**

A method applicable to fill gaps and repair data with gaps has been developed. Three different versions of the method – from simplest linear models to the method of principal curves for the data with gaps have been presented. The neural network implementation of the method makes possible easy construction of its parallel implementations.

The given algorithm of filling gaps - as opposed to many other algorithms designed for this purpose - does not involve its a priori filling. However, it calls for preliminary normalization ("dedimensionalizing") of the data – transition in each column of the table to a "natural" unit of measurement. It is noteworthy that data centering cannot turn the problem of processing data with gaps into a homogeneous problem.

Of great interest is the question: how many summands (principal curves) should be taken to process the data? There are several versions of answers, yet most of them are subject to the heuristic formula: *the number of summands must be minimal among those that provide for satisfactory (tolerant) testing of the method with the known data*. Such a principle of "minimum sufficiency" is specific for many neural network applications [8-10].

The method developed manifests itself in the form of an "anzatz" – a sentence, but not a series of theorems. This is not incidental – we propose a technology of constructing *plausible* evaluations of missing data, but not of their unknown real value. The practical value of the methods of this kind should be assessed and weighted by the users of this technology. Appropriate software has been developed. It showed good performance in solving difficult problems with a large number of missing data, and in simpler (standard) cases yields results identical to classical methods of statistical analysis.